\documentclass[letterpaper,11pt]{article}

\usepackage{diagrams}
\usepackage{enumerate}
\usepackage{psfrag}
\usepackage{pstricks,pst-node}
\usepackage{hyperref}

\usepackage[dvips]{graphicx}
\usepackage[usenames]{color}
\usepackage{psfrag}

\usepackage{latexsym}    
\usepackage{amsmath}
\usepackage{amssymb}
\usepackage{amsthm}
\usepackage{slashed}
\usepackage{xy}
\xyoption{all}

\theoremstyle{plain}
\newtheorem{theorem}{Theorem}[section]
\newtheorem{lemma}[theorem]{Lemma}

\newtheorem{corollary}[theorem]{Corollary}

\theoremstyle{definition}
\newtheorem{defn}[theorem]{Definition}
\newtheorem{remark}[theorem]{Remark}
\newtheorem{example}[theorem]{Example}

\renewcommand{\dag}{\ensuremath{\dagger}}

\newcommand\ignore[1]{}

\newcommand\jskip{3pt}

\newcommand{\cat}[1]{\ensuremath{\mathbf{#1}}}
\newcommand{\id}{\ensuremath{ \mathrm{id} }}

\newcommand{\jbeginproof}{\begin{proof}\vspace{-\jskip}}
\newcommand{\jbeginenumerate}{\begin{enumerate}\vspace{-\jskip}}

\newcommand{\ket}[1]{\ensuremath{| #1 \rangle}}
\newcommand{\bra}[1]{\ensuremath{\langle #1 |}}
\newcommand{\braket}[2]{\ensuremath{\langle #1 | #2 \rangle}}

\newcommand{\vc}[1]
{
\begin{array}{c}
#1
\end{array}
}

\newcommand{\bit}{\begin{itemize}}
\newcommand{\eit}{\end{itemize}\par\noindent}
\newcommand{\ben}{\begin{enumerate}}
\newcommand{\een}{\end{enumerate}\par\noindent}
\newcommand{\beq}{\begin{equation}}
\newcommand{\eeq}{\end{equation}\par\noindent}
\newcommand{\beqa}{\begin{eqnarray*}}
\newcommand{\eeqa}{\end{eqnarray*}\par\noindent} 
\newcommand{\beqn}{\begin{eqnarray}}
\newcommand{\eeqn}{\end{eqnarray}\par\noindent}

\newcommand\I{I}   
\newcommand{\mult}{m}
\newcommand{\unit}{u}

\begin{document}

\title{\bf A new description of orthogonal bases}

\author{Bob Coecke, Dusko Pavlovic and Jamie Vicary\\ Oxford University Computing Laboratory}

\date{}
\maketitle
\begin{abstract}
We show that an orthogonal basis for a finite-dimensional Hilbert space can be equivalently characterised as a commutative $\dagger$-Frobenius monoid in the category \cat{FdHilb}, which has finite-dimensional Hilbert spaces as objects and continuous linear maps as morphisms, and tensor product
for the monoidal structure.  The basis is
normalised exactly when the corresponding
 commutative $\dagger$-Frobenius monoid is special.  Hence orthogonal and orthonormal bases can be axiomatised in terms of composition of operations and tensor product only, without any explicit reference to the underlying vector spaces.  This axiomatisation moreover admits an operational interpretation, as
the comultiplication copies the basis vectors
and the counit uniformly deletes them.  That is, we rely on the distinct ability to clone and delete classical data as compared to quantum data to capture basis vectors.  For this reason our result has important implications for categorical quantum mechanics.
\end{abstract}

\section{Introduction}

Given any orthonormal basis $\{|\phi_i\rangle\}_i$ in a finite dimensional Hilbert space $H$ 
we can always define the linear maps
\beq\label{eq:copy}
\delta: H\to H\otimes H:: |\phi_i\rangle\mapsto |\phi_i\rangle\otimes|\phi_i\rangle 
\eeq
and 
\beq\label{eq:delete}
\epsilon: H\to\mathbb{C}::|\phi_i\rangle\mapsto 1\,.
\eeq
It was observed in \cite{Coecke-Pavlovic} that the triple $(H,\delta,\epsilon)$ is a so-called \em commutative special $\dag$-Frobenius comonoid \em in the category \cat{FdHilb} of finite dimensional Hilbert spaces and linear maps with the tensor product as monoidal structure.  
Meanwhile, this fact that orthonormal bases can be encoded as commutative special $\dag$-Frobenius monoids has resulted in many important applications in the area of categorical quantum mechanics, for example,  for describing the flow of classical information in quantum informatic protocols \cite{Coecke-Pavlovic},  for defining complementarity and special quantum logic gates in quantum computational schemes \cite{Coecke-Duncan} and for constructing discrete models for quantum reasoning \cite{Spek}.  In this paper, we establish that every commutative special $\dag$-Frobenius monoids arises from an orthonormal  basis, and that dropping the specialty condition gives an orthogonal basis.


The plan of the paper is as follows:
\begin{itemize}
\item 
Section \ref{sec:prelem} provides category-theoretic preliminaries.
\item 
Section \ref{sec:BaseToMon} spells out in more detail how a commutative \dag-Frobenius monoid arises from an orthogonal basis.
\item 
Section \ref{sec:MonToBase}  describes how to extract an orthogonal  basis from any  commutative \dag-Frobenius monoid in \cat{FdHilb}. 
\item 
Section \ref{sec:Result} states the main theorem.
\item 
Section \ref{sec:other} spells out that within this established bijective correspondence normalisation of basis vectors means speciality of the corresponding  commutative \dag-Frobenius monoid.  We also compare this result to an already-known result which classifies \emph{arbitrary} bases on a finite-dimensional complex vector space as special Frobenius algebras.
\item Section \ref{sec:Categories} describes some categorical statements that come out of these results; in particular, we describe how to obtain the category of finite sets as a category of commutative \dag-Frobenius monoids in \cat{FdHilb}.
\end{itemize}

\section{Preliminaries}\label{sec:prelem}

The research area of categorical quantum mechanics emerged from the observation that the subtle details of important, experimentally-established quantum informatic protocols can already be specified at an abstract category-theoretic level \cite{Abramsky-Coecke}.  The background structure is that of a symmetric monoidal $\dagger$-category  \cite{Abramsky-Coecke, Selinger}, a symmetric monoidal category together with a identity-on-objects involutive endofunctor which coherently preserves the symmetric monoidal structure. Within this context one then aims to maximise the 
\[
{\mbox{expressiveness}\over\mbox{additional structure}}
\]
ratio.  Additional structure on which we rely in this paper is that of internal Frobenius algebras \cite{CarboniWalters, LawvereFrobenius}, more specifically, internal commutative $\dagger$-Frobenius monoids \cite{Coecke-Pavlovic}.  Relative to the \em quantum universe \em which is modelled by the symmetric monoidal $\dagger$-category these commutative $\dagger$-Frobenius monoids model the   \em classical interfaces\em, and enable us to specify projector spectra, measurements, and classical data flows \cite{Coecke-Pavlovic}.

\begin{defn}
\label{def:frobmon}
A {\em Frobenius monoid} in a  symmetric monoidal category
is a quintuple $(X, m,u,\delta, \epsilon)$ consisting of an internal monoid 
\[
\xymatrix@=0.44in{
\I\ar[r]^{\unit} &X& \ar[l]_{\mult\ \ } X\otimes X
}
\] 
and an internal comonoid 
\[
\xymatrix@=0.44in{
\I & \ar[l]_{\epsilon}X\ar[r]^{\delta} &  X\otimes X
}
\]
which together satisfy the {\em Frobenius condition\/} 
\[
\xymatrix{
X\otimes X \ar[dr]^-{\mult} \ar[dd]_-{\delta\otimes X} \ar[rr]^{X\otimes \delta}&& X\otimes X\otimes X \ar[dd]^-{\mult\otimes X} \\
& X \ar[dr]^-{\delta} \\
X\otimes X \otimes X \ar[rr]_-{X\otimes \mult} && X\otimes X
}
\]
A Frobenius algebra is called\/ {\em special}  if
\[
\ \mult\circ \delta = \id_X,
\]
and it is \em commutative \em if 
\[
\ \sigma \circ \delta = \delta.
\]
A\/ {\em (special) (commutative) $\dagger$-Frobenius monoid} in a  symmetric monoidal $\dagger$-category is a triple $(X,\mult,\unit)$ such that  $(X,\mult,\unit,\delta=\mult^\dagger,\epsilon=\unit^\dagger)$ is a (special) (commutative) Frobenius algebra.
\end{defn}

\begin{remark}
We will also use $\dagger$-Frobenius \em co\em monoid to refer to a $\dagger$-Frobenius monoid, depending on whether we want the emphasis to lie either on the monoid or the comonoid structure.
\end{remark}

\begin{example}
The monoidal unit $\I$ comes with a canonical special commutative $\dagger$-Frobenius comonoid,
namely  $(\I,\lambda_{ \I}:{ \I}\simeq { I}\otimes { \I},\id _{ \I})$.
\end{example}

Recall that in a symmetric monoidal category a morphism $f:X\to Y$ is a \em monoid homomorphism \em for monoids $(X,m,u)$ and $(Y,m',u')$ if 
\[
f\circ m=m'\circ (f\otimes f)\quad \mbox{and} \quad f\circ u=u',
\]
and that it is a comonoid homomorphism for comonoids $(X,\delta,\epsilon)$ and $(Y,\delta',\epsilon')$ if
\[
\delta\circ f= (f\otimes f)\circ\delta'\quad \mbox{and} \quad \epsilon \circ f =\epsilon'\,.
\]

\begin{defn} A \em copyable element \em of a $\dagger$-Frobenius monoid  $(X,m_X,u_X)$ is a comonoid homomorphism $\alpha: \I \to X$.
\end{defn}

\section{Turning an orthogonal basis into a commutative \dag-Frobenius monoid}\label{sec:BaseToMon} 

Given an orthogonal basis $\{|\phi_i\rangle\}_i$, the maps defined in (\ref{eq:copy}) and (\ref{eq:delete}) are the linear extensions of \em copying \em and \em uniformly deleting \em these basis vectors.   

It is easily seen that from $\delta$ alone we can recover the basis  by solving
\[
\delta(|\psi\rangle)=|\psi\rangle\otimes |\psi\rangle\,.
\]
No other vectors besides those in $\{|\phi_i\rangle\}_i$ will satisfy this equation since for 
\[
|\psi\rangle=\sum_{i=1} ^{\textrm{dim}(H)} \frac{\braket{\phi_i}{\psi}} {\braket{\phi_i}{\phi_i}}\,|\phi_i\rangle
\]
with at least two non-zero scalars in $\{\braket{\phi_i}{\psi}\}_i$ we have that 
\[
\delta(|\psi\rangle)=\sum_{i=1} ^{\textrm{dim}(H)} \frac{\braket{\phi_i}{\psi}} {\braket{\phi_i}{\phi_i}}\,(|\phi_i\rangle\otimes|\phi_i\rangle)
\]
will always be \em entangled\em, i.e.~cannot be written in the form $|\psi_a\rangle\otimes |\psi_b\rangle$ for some $|\psi_a\rangle$ and $|\psi_b\rangle$, and hence not equal to $|\psi\rangle\otimes|\psi\rangle$. So $\delta$ and hence also the triple $(H,\delta,\epsilon)$ faithfully  encodes $\{|\phi_i\rangle\}_i$.  

To see that $\delta^\dagger$ and $\delta$ obey the Frobenius condition it suffices to note that 
\beq\label{eq:copyDAG}
\delta^\dag: H\otimes H\to H:: |\phi_i\rangle\otimes|\phi_j\rangle \mapsto 
\left\{\begin{array}{cl}
|\phi_i\rangle& i=j\\
0& i\not=j
\end{array}\right.
\eeq
so
\[
|\phi_i\rangle\otimes|\phi_j\rangle
\stackrel{\delta^\dag}{\mapsto}
\left\{\begin{array}{c}
|\phi_i\rangle\\
0
\end{array}\right.
\stackrel{\delta}{\mapsto}
\left\{\begin{array}{cl}
|\phi_i\rangle\otimes |\phi_i\rangle& i=j\\
0& i\not=j
\end{array}\right.
\]
and 
\[
|\phi_i\rangle\otimes|\phi_j\rangle
\stackrel{\id_H\otimes \delta}{\mapsto}
|\phi_i\rangle\otimes|\phi_j\rangle\otimes|\phi_j\rangle
\stackrel{\delta^\dagger\otimes\id_H}{\mapsto}
\left\{\begin{array}{cl}
|\phi_i\rangle\otimes |\phi_i\rangle& i=j\\
0& i\not=j
\end{array}\right.\,.
\]
As a consequence, by linearity, $\delta\circ\delta^\dag=(\delta^\dagger\otimes\id_H)\circ(\id_H\otimes \delta)$.  That 
$(X, \delta,\epsilon)$ is a comonoid is easily verified.  The unit of the corresponding monoid is 
\beq\label{eq:deleteDAG}
\epsilon^\dagger: \mathbb{C}\to H::1\mapsto \sum_{i=1} ^{\textrm{dim}(H)} \!\!\! |\phi_i\rangle.
\eeq

\section{Turning a commutative \dag-Frobenius monoid into an orthogonal basis}\label{sec:MonToBase}

We will freely switch between denoting elements of $H$ as linear maps 
\[
\alpha : \mathbb{C} \to H::1\mapsto \ket{\alpha}
\] 
and as kets $\ket{\alpha}=\alpha(1) \in H$. Taking the adjoint of $\alpha$ gives us 
\[
\alpha  ^\dag : H \to \mathbb{C}::|\psi\rangle\mapsto \braket{\alpha}{\psi}
\]
and hence $\bra{\alpha}=\alpha  ^\dag \in H^*$. 


Let $(H,m,u)$ be a commutative \dag-Frobenius monoid. Given such a commutative \dag-Frobenius monoid any element $\alpha \in H$ induces a linear map 
\[
R _\alpha := m \circ (\id _A \otimes \alpha) : H \to H\,,
\]
its \em its right action\em. We draw this right action in the following way:
\[
\psfrag{p}{\hspace{0pt}$\alpha$}
\includegraphics[scale=1]{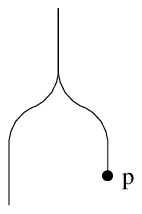}
\]
The diagram is read from bottom to top. This is a direct representation of our definition of $R _\alpha$ as right-multiplication by the element $\alpha$: vertical lines represent the vector space $H$, the dot represents the element $\alpha$, and the merging of the two lines represents the multiplication operation $m$. Since $H$ is a Hilbert space, $R _\alpha$ has an adjoint
\[
R _\alpha {}^\dag : H \to H.
\] 
We draw the adjoint $R _\alpha {}^\dag$ by flipping the diagram on a horizonal axis, but keeping the arrows pointing in their original direction:
\[
\psfrag{d}{$\alpha ^\dag$}
\includegraphics{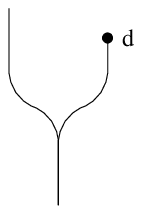}
\]
The splitting of the line into two represents the comultiplication, and the dot represents the linear map $\alpha ^\dag : H \to \mathbb{C}$.  We show that this adjoint to a right action is also the right action of some element of $H$.

\begin{lemma}\label{lm:conjugate}
If $(X,m,u)$ is a commutative \dag-Frobenius monoid in a symmetric monoidal \mbox{$\dagger$-category} then
$R _{\alpha} {}^\dag = R _{\alpha'}$
for
$\alpha ' = (\id _{X} \otimes \alpha ^\dag) \circ m^\dag \circ u$.
\end{lemma}
\begin{proof}
We draw the Frobenius law of  definition \ref{def:frobmon} in the following way:
\[
\vc{\includegraphics{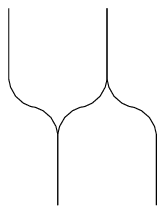}}
=
\vc{\includegraphics{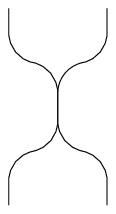}}
=
\vc{\includegraphics{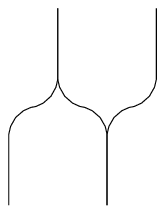}}
\]
Representing the unit $u: \I \to X$ as a small horizontal bar we draw the  unit law in the following way:
\[
\vc{\includegraphics{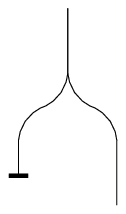}}
=
\vc{\includegraphics{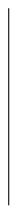}}
=
\vc{\includegraphics{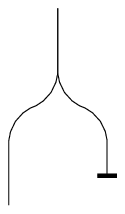}}
\]
We can now use the unit law and the Frobenius law to redraw the graphical representation of $R _\alpha {}^\dag$ in the following way:
\[
{
\hspace{-15pt}
\psfrag{d}{$\alpha ^\dag$}
\vc{\includegraphics{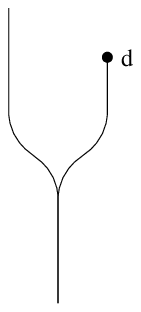}}
\hspace{-15pt}
}
=
{
\hspace{-15pt}
\psfrag{d}{$\alpha ^\dag$}
\vc{\includegraphics{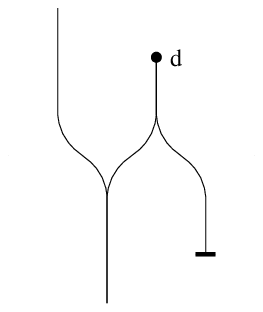}}
\hspace{-15pt}
}
=
{
\hspace{-15pt}
\psfrag{p}{$\alpha ^\dag$}
\vc{\includegraphics{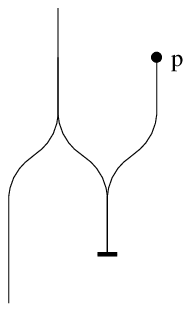}}
\hspace{-15pt}
}
=
{
\hspace{-15pt}
\psfrag{s}{$\alpha ^\dag$}
\vc{\includegraphics{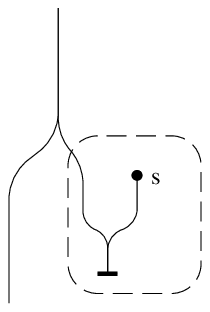}}
\hspace{-15pt}
}
\equiv{
\hspace{-15pt}
\psfrag{ap}{$\alpha '$}
\vc{\includegraphics{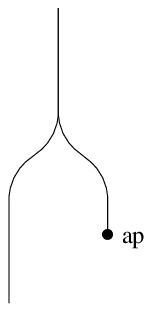}}
\hspace{-15pt}
}
\]
So the adjoint of $R _\alpha$ is indeed itself a right action of $\alpha'$, as defined above.
\end{proof}


\begin{lemma}
Let ${\bf C}$ be a symmetric monoidal $\dagger$-category and $(X,m,u)$ a commutative \mbox{\dag-Frobenius} monoid in  ${\bf C}$. The right action mapping 
\[
{\bf C}({\rm I}, X)\to {\bf C}(X,X)::\alpha\mapsto R_\alpha
\]
is an involution preserving monoid embedding, when endowing ${\bf C}({\rm I}, X)$ with the monoid structure of the internal monoid $(X,m,u)$. In the case that ${\bf C}=\cat{FdHilb}$ then this mapping also preserves the vector space structure.
\end{lemma}
\begin{proof}
Using the Frobenius and unit identities and the fact that the \dag-functor is an involution we first show that $(-)'$ as in Lemma \ref{lm:conjugate} is involutive:
\[
\vspace{-10pt}
\psfrag{a}{\hspace{11pt}$(\alpha')'$}
{
\hspace{-20pt}
\vc{\includegraphics{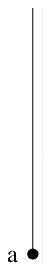}}
\hspace{-5pt}
}
=
{
\hspace{-5pt}
\psfrag{a}{\hspace{-0pt}$(\alpha' ) ^\dag$}
\vc{\includegraphics{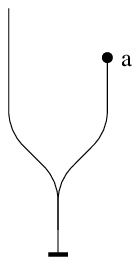}}
\hspace{-5pt}
}
=
{
\hspace{-5pt}
\psfrag{a}{\hspace{-0pt}$(\alpha ^\dag) ^\dag$}
\vc{\includegraphics{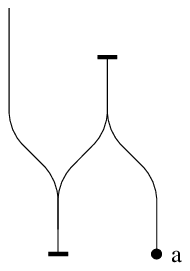}}
\hspace{-5pt}
}
=
{
\hspace{-5pt}
\psfrag{a}{\hspace{11pt}$\alpha$}
\vc{\includegraphics{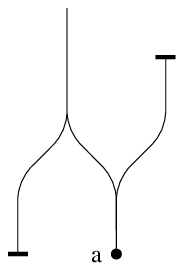}}
\hspace{-5pt}
}
=
\psfrag{a}{\hspace{11pt}$\alpha$}
{
\hspace{-5pt}
\vc{\includegraphics{graphics/newalphainv-3}}
\hspace{-15pt}
}
\]
The adjoint  is an involution on ${\bf C}(X,X)$ so since $R _{\alpha'}=R _{\alpha} {}^\dag$ involution is indeed preserved.  The mapping  is moreover injective, since by the unit equation we have $R _{\alpha}\circ u= \alpha$.  It is straightforward to show that multiplication and unit are preserved. Preservation of the vector space structure in the case that ${\bf C}=\cat{FdHilb}$ follows by linearity of $m$.
 \end{proof}
 
\begin{corollary}
Any \dag-Frobenius monoid in \cat{FdHilb} is a C*-algebra.
\end{corollary}
\begin{proof}
The endomorphism monoid $\cat{FdHilb}(H,H)$ is a C*-algebra.  By the above Lemma we know that
\[
H\simeq \cat{FdHilb}(\mathbb{C},H)\simeq R_{[\cat{FdHilb}(\mathbb{C},H)]}\subseteq\cat{FdHilb}(H,H)
\]
inherits algebra structure from $\cat{FdHilb}(H,H)$.  Now, since any finite-dimensional involution-closed subalgebra of a C*-algebra is also a C*-algebra, it follows that any \dag-Frobenius monoid in \cat{FdHilb} is a C*-algebra, in particular, it can be given a C*-algebra norm.
\end{proof}

\begin{remark}
Note that in the above we did not assume the \dag-Frobenius monoid to be commutative. More on this is in  \cite{Jamie}.
\end{remark}

\begin{corollary}\label{col:spectral}
The copyable elements for any commutative \dag-Frobenius monoid on $H$ in \cat{FdHilb} form a basis for $H$.  
\end{corollary}

\begin{proof}
By the spectral theorem for finite-dimensional commutative \mbox{C*-algebras} \cite{m90-caot}, the involution-preserving homomorphisms from a finite-dimensional commutative C*-algebra to the complex numbers form a basis for the dual of the underlying vector space, in our case $H\simeq \cat{FdHilb}(\mathbb{C},H)\simeq R_{[\cat{FdHilb}(\mathbb{C},H)]}$.  We write these states as ${\phi_i ^\dag} : H \to \mathbb{C}$, and in our case they form a basis for $H^*$. Their adjoints $\phi_i : \mathbb{C} \to H$ will therefore form a basis for $H$. Since the morphisms $\phi_i ^\dag : H \to \mathbb{C}$ are monoid homomorphisms from the monoid $(H,m,u)$ to the monoid $(\mathbb{C},\lambda_\mathbb{C},\id _\mathbb{C})$, by taking adjoints we see that the morphisms $\phi_i : \mathbb{C} \to H$ are comonoid homomorphisms from the comonoid $(\mathbb{C},\lambda_\mathbb{C},\id _\mathbb{C})$ to the comonoid $(H,\delta = m ^\dag, \epsilon =u ^\dag)$.

To make use of the spectral theorem, we also need to show that these homomorphisms are involution-preserving. In fact, this is automatic: if a map between two finite-dimensional commutative C*-algebras preserves the algebra multiplication and unit, then it necessarily preserves the involution.
\end{proof}

\newcommand\first{\phi_i}
\newcommand\second{\phi _j}

It remains to be shown that this basis is orthogonal.

\begin{lemma}\label{lm:cancel}
If $\phi_i,\phi_j: \I \to X$ are comonoid homomorphisms for a commutative \dag-Frobenius comonoid and if $\braket{\phi_i}{\phi_j}:=\phi_i^\dagger\circ\phi_j$, $\braket{\phi_i}{\phi_i}$ and  $\braket{\phi_j}{\phi_j}$ are all cancellable scalars then they must be real and equal.
\end{lemma}

\begin{proof}
Given that $\first$ and $\second$ are comonoid homomorphisms, and making use of one of the Frobenius identities, we can derive the following equation:
\[
\psfrag{f}{$\first$}
\psfrag{p}{$\second$}
\psfrag{fd}{$\first ^\dag$}
\psfrag{pd}{$\second ^\dag$}
{
\hspace{-0pt}
\vc{\includegraphics{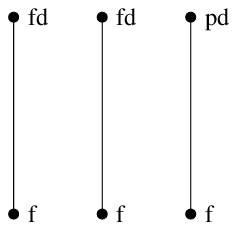}}
\hspace{-10pt}
}
=
{
\hspace{-0pt}
\vc{\includegraphics{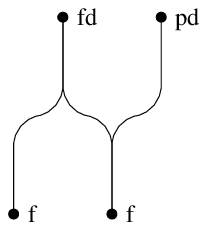}}
\hspace{-10pt}
}
=
{
\hspace{-0pt}
\vc{\includegraphics{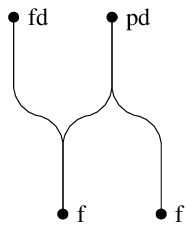}}
\hspace{-10pt}
}
=
{
\hspace{3pt}
\vc{\includegraphics{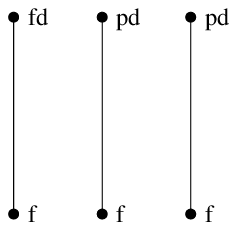}}
\hspace{-10pt}
}
\]
Switching the roles of $\first$ and $\second$ we can obtain another similar equation, and writing both equations in bra-ket notation we obtain
\[
\braket{\first}{\first}
\braket{\first}{\first}
\braket{\first}{\second}
=
\braket{\first}{\first}
\braket{\first}{\second}
\braket{\first}{\second},
\]
\[
\braket{\second}{\second}
\braket{\second}{\second}
\braket{\second}{\first}
=
\braket{\second}{\second}
\braket{\second}{\first}
\braket{\second}{\first}.
\]
If we cancel  $\braket{\first}{\first}$, $\braket{\second}{\second}$ and $\braket{\first}{\second}$ we obtain
\[
\braket{\first}{\first} = \braket{\first}{\second} 
\qquad\mbox{and}\qquad
\braket{\second}{\second} =  \braket{\second}{\first}.
\]
It follows that these inner products are real, and that they are all equal:
\[
\braket{\first}{\first} = \braket{\first}{\second} = \braket{\second}{\first} = \braket{\second}{\second}.
\qedhere
\]
\end{proof}

\begin{corollary}
The copyable elements for any commutative \dag-Frobenius monoid on $H$ in \cat{FdHilb} form an orthogonal basis for $H$.  
\end{corollary}
\begin{proof}
Let $\first,\second : \mathbb{C} \to H$ be two different vectors in this basis. This will only be impossible to satisfy when $H$ is one-dimensional, but in that case the basis is trivially orthogonal. 
Assume that $\first$ and $\second$ are \textit{not} orthogonal, i.e.~$\braket{\first}{\second}$ is cancellable, so Lemma \ref{lm:cancel} applies.
Since
\[
\braket{\first-\second}{\first-\second}
= \braket{\first}{\first} - \braket{\first}{\second}  - \braket{\second}{\first} + \braket{\second}{\second}
 = 0
\]
we have $\ket{\first} - \ket{\second} = 0$, so $\ket{\first}$ and $\ket{\second}$ are the same. This contradicts our assumption that ${\first}$ and ${\second}$ were different, and so the basis $\ket{\phi_i}$ is orthogonal.
\end{proof}

\section{Statement of the main result}\label{sec:Result}

\begin{theorem}
Every commutative \dag-Frobenius monoid in \cat{FdHilb} determines an orthogonal basis, consisting of its copyable elements, and every orthogonal basis determines a commutative \dag-Frobenius monoid in \cat{FdHilb} via prescriptions (\ref{eq:copy}) and (\ref{eq:delete}).  These constructions are inverse to each other.
\end{theorem}

It only remains to be explained why the construction described in section \ref{sec:MonToBase}, which obtains an orthogonal basis from a commutative \dag-Frobenius algebra, is inverse to the construction described in section \ref{sec:BaseToMon}, which obtains a commutative \dag-Frobenius algebra from an orthogonal basis.

Assume we begin with an orthogonal basis. We construct the \dag-Frobenius monoid as the unique one with a comultiplication which perfectly copies our original basis, and we construct a new orthogonal basis consisting of those elements which are perfectly copied by the comultiplication. This new basis must be at least as large as our original basis. However, since any two bases for a finite-dimensional Hilbert space must have the same number of elements, the new basis is actually the same as the original basis.

Now assume we begin with a commutative \dag-Frobenius monoid. We construct our orthogonal basis as those elements which are perfectly copied, and construct our new monoid as the unique \dag-Frobenius monoid which perfectly copies this basis. However, since the original monoid also perfectly copies this basis, by uniqueness the two monoids are the same.

\begin{remark}
In other papers on this subject abstract basis vectors are required to be \em self-conjugate \em comonoid homomorphisms, where the conjugate of a morphism $f:X\to Y$ relative to $\dagger$-Frobenius monoids $(X,m_X,u_X)$ and $(Y,m_Y,u_Y)$ is defined to be 
\[
f_*:= (\id_Y\otimes\eta_X{}^\dagger)\circ(\id_Y\otimes f^\dagger\otimes\id_X) \circ(\eta_Y\otimes\id_X)
\]
with $\eta_Z=m_Z{}^\dagger\circ u_Z:{\rm I}\to Z\otimes Z$.  In \cat{FdHilb} a morphism is self-conjugate if in its matrix representation in the corresponding bases all entries are self-conjugate, a fact which follows in \cat{FdHilb} automatically from the fact of being a comonoid homomorphism. In other categories the additional constraint of the basis vectors being self-conjugate guarantees that they are involution preserving -- cf.~the last part of the proof of corollary \ref{col:spectral}.
\end{remark}

\section{Other types of basis}
\label{sec:other}

The elements of the orthogonal basis are normalised exactly when the corresponding commutative \mbox{\dag-Frobenius} algebra is \emph{special}, meaning \mbox{$m \circ \delta = \id _H$}. This is most straightforwardly seen from the explicit form of the comultiplication in terms of the orthogonal basis as given in (\ref{eq:copy}). From this we see that
\[
m \circ \delta = \delta ^\dag \circ \delta = \sum _{i=1} ^{\mathrm{dim}(H)} \ket{\phi_i} \bra{\phi _i},
\]
and it is clear that this is the identity if and only if the vectors $\ket{\phi_i}$ are normalised.

Interestingly, it is already known that, for a finite-dimensional complex \emph{vector} space, a basis exactly corresponds to a choice of special commutative Frobenius algebra\footnote{Thanks to John Baez for pointing this out.}. Of course, it does not make sense to ask whether such a Frobenius algebra is \dag-Frobenius, or whether the elements of such a basis are normalised or orthogonal. This result follows from the fact that a special commutative Frobenius algebra is necessarily strongly separable, and since the ground field is of characteristic~0, it is therefore necessarily finite-dimensional and semisimple \cite{Aguiar}. Such an algebra is canonically isomorphic to a finite cartesian product of the complex numbers, up to permutation, and the basis elements are given by the number $1$ in each of the complex factors.

In summary, on a finite-dimensional complex Hilbert space, we can describe different types of basis very precisely with the following structures:
\begin{center}
\begin{tabular}{l@{\hspace{30pt}}l}
\bf
Type of basis
& \bf
Algebraic structure
\\
Arbitrary
&
Commutative special Frobenius algebra
\\
Orthogonal
&
Commutative \dag-Frobenius algebra
\\
Orthonormal
&
Commutative special \dag-Frobenius algebra
\end{tabular}
\end{center}
The inner product on the Hilbert space does not play a role for the case of the arbitrary basis. In every case, the basis is recovered from the Frobenius algebra as those vectors which are perfectly copied by the comultiplication, and the Frobenius algebra is recovered from the basis as the unique Frobenius algebra of the correct type with a comultiplication that perfectly copies the basis.

It is interesting to consider whether an arbitrary commutative Frobenius algebra on a complex vector space might also correspond to some type of basis structure. In fact, it does not, and such Frobenius algebras can be very wild indeed. It is perhaps surprising that the specialness axiom and the \dag-Frobenius axiom can both serve independently to tame this wildness.

\section{Categorical statements}\label{sec:Categories}

We have shown that a commutative \dag-Frobenius monoid in \cat{FdHilb} corresponds to a basis of a finite-dimensional Hilbert space. Any such basis is determined up to unitary isomorphism by the norms of the basis elements, which constitute a list of positive real numbers.

If a homomorphism between two commutative \dag-Frobenius monoids preserves all of the structure --- the multiplication, unit, comultiplication and counit --- then it is necessarily an isomorphism, and in \cat{FdHilb}, it is necessarily unitary. Such a homomorphism will map one basis onto another, taking basis elements onto basis elements of the same norm. This leads to the following result:

\begin{corollary}
The category of commutative \dag-Frobenius monoids in \cat{FdHilb}, with morphisms preserving all of the Frobenius structure,  is equivalent to the groupoid of `finite lists of real numbers and isomorphisms that preserve the numbers', which has objects given by finite sets equipped with functions into the positive real numbers, and morphisms given by isomorphisms of sets that preserve the functions into the real numbers.
\end{corollary}
\noindent
This is interesting from the perspective of unitary 2-dimensional topological quantum field theory \cite{k04-fa2d}, since such things are given by commutative \mbox{\dag-Frobenius} monoids in \cat{FdHilb}, and the natural notion of homomorphism is one that preserves all of the Frobenius structure.  

If we only require that our homomorphisms preserve the
comultiplication and counit then this gives arbitrary functions
between bases, without the requirement of preserving the length of the
basis vector. This follows from the spectral theorem for commutative
C*-algebras; a comonoid homomorphism gives rise to an
oppositely-directed monoid homomorphism by taking the adjoint, and any
involution-preserving monoid homomorphism is equivalent to an
oppositely-directed continuous function between the spectra of the
C*-algebras.

\begin{corollary}
\label{finsetlemma}
The category of commutative \dag-Frobenius monoids in \cat{FdHilb},
with morphisms preserving comultiplication and counit, is equivalent
to the category \cat{FinSet} of finite sets and functions.
\end{corollary}
\noindent
This gives an interesting new perspective on the relationship between
\cat{FdHilb} and \cat{FinSet}. There is an obvious functor $F:
\cat{FinSet} \to \cat{FdHilb}$ which takes a set with $n$ elements to
a Hilbert space of dimension $n$ and chosen basis, and takes a
function between sets to the induced linear map. This can be thought
of as a `free' functor, as it generates the free Hilbert space on a
finite set. But using the equivalence between \cat{FinSet} and the
category of commutative \dag-Frobenius monoids described in lemma
\ref{finsetlemma}, we see that this functor $F$ is equivalent to the
\emph{forgetful} functor which regards each finite set as a Hilbert
space equipped with a commutative \mbox{\dag-Frobenius} monoid
structure, and forgets the monoid structure. So a finite set can be
considered as a finite-dimensional Hilbert space with the extra
structure of a commutative \dag-Frobenius monoid, or a
finite-dimensional Hilbert space can be considered as a finite set
with the extra structure of a vector space and inner product.

\end{document}